\documentclass[conference]{IEEEtran}
\IEEEoverridecommandlockouts

\usepackage{cite}
\usepackage{amsmath,amssymb,amsfonts}
\usepackage{algorithmic}
\usepackage{graphicx}
\usepackage{textcomp}
\usepackage{xcolor}
\usepackage{tabularx}
\usepackage{hyperref}
\def\BibTeX{{\rm B\kern-.05em{\sc i\kern-.025em b}\kern-.08em
    T\kern-.1667em\lower.7ex\hbox{E}\kern-.125emX}}
\begin{document}

\title{Speech-Declipping Transformer with Complex Spectrogram and Learnerble Temporal Features\\
\thanks{This work was supported by the National Research Foundation of Korea (NRF) grant (No. RS-2024-00337945) and STEAM research grant (No. RS-2024-00337945) funded by the Ministry of Science and ICT of Korea government (MSIT), the BK21 FOUR program through the NRF grant funded by the Ministry of Education of Korea government (MOE).}
}

\author{\IEEEauthorblockN{1\textsuperscript{st} Younghoo Kwon}
\IEEEauthorblockA{\textit{School of Electrical Engineering} \\
\textit{KAIST}\\
Daejeon, South Korea \\
k0hoo@kaist.ac.kr}
\and
\IEEEauthorblockN{2\textsuperscript{nd} Jung-Woo Choi}
\IEEEauthorblockA{\textit{School of Electrical Engineering} \\
\textit{KAIST}\\
Daejeon, South Korea \\
jwoo@kaist.ac.kr}
}

\maketitle

\begin{abstract}
We present a transformer-based speech-declipping model that effectively recovers clipped signals across a wide range of input signal-to-distortion ratios (SDRs).
While recent time-domain deep neural network (DNN)-based declippers have outperformed traditional handcrafted and spectrogram-based DNN approaches, they still struggle with low-SDR inputs.
To address this, we incorporate a transformer-based architecture that operates in the time-frequency (TF) domain. The TF-transformer architecture has demonstrated remarkable performance in the speech enhancement task for low-SDR signals but cannot be optimal for the time-domain artifact like clipping.   
To overcome the limitations of spectrogram-based DNNs, we design an extra convolutional block that directly extracts temporal features from time-domain waveforms. The joint analysis of complex spectrogram and learned temporal features allows the model to improve performance on both high- and low-SDR inputs.
Our approach also preserves the unclipped portions of the speech signal during processing, preventing degradation typically seen when only spectral information is used.
In evaluations on the VoiceBank-DEMAND and DNS challenge datasets, the proposed model consistently outperformed state-of-the-art (SOTA) declipping models across various metrics, demonstrating its robustness and generalizability.
\end{abstract}


\section{Introduction}
Speech clipping refers to limiting speech waveform's amplitudes exceeding a certain threshold and often occurs due to the limited dynamic range of a recording system. Clipping can significantly degrade the perceived quality and intelligibility of speech 
through spectral distortions, 
such as the introduction of harmonic distortions in the high-frequency region~\cite{laguna2016efficient, perception_analysis, AliasingDistortion}. Such distortions also adversely affect the performance of automatic speech recognition (ASR) systems~\cite{harvilla2014least}. 

Speech declipping (SD) techniques have been developed to reconstruct the original speech from the clipped waveform.
Classical sparse-optimization-based methods~\cite{sparse_2013, sparse_2015_aspade} introduce a constraint to preserve the unclipped portions of a waveform, referred to as reliable samples. These methods perform well with signals containing a large number of reliable samples, such as those with high clipping thresholds or high SDRs.
One notable approach, A-SPADE~\cite{sparse_2015_aspade}, has shown solid reconstruction performance across a variety of declipping metrics~\cite{servey_article}.

Recently, DNN-based declipping approaches~\cite{deepfiltering, tunet} have demonstrated superior performance on low SDR inputs. For example, deep filtering approach~\cite{deepfiltering} utilized bidirectional long short-term memory (BLSTM) to recover complex short-time Fourier transform (STFT) data. T-UNet~\cite{tunet} employed a U-Net\cite{unet} architecture and sub-pixel convolutions to realize declipping in time domain. However, these methods do not explicitly leverage reliable samples, which often results in degraded performance on high-SDR inputs, thus limiting the generalizability of the models. Hybrid methods such as Applade~\cite{applade} and Upglade~\cite{upglade} were introduced, utilizing consensus equilibrium by applying A-SPADE for high-SDR inputs and DNNs for low-SDR inputs.
While these hybrid methods have shown improved performance, the improvement by DNNs for low-SDR inputs remains limited.

\begin{figure*}[ht]
    \centering
    \vspace{-1em}
    \includegraphics[width=0.9\textwidth]{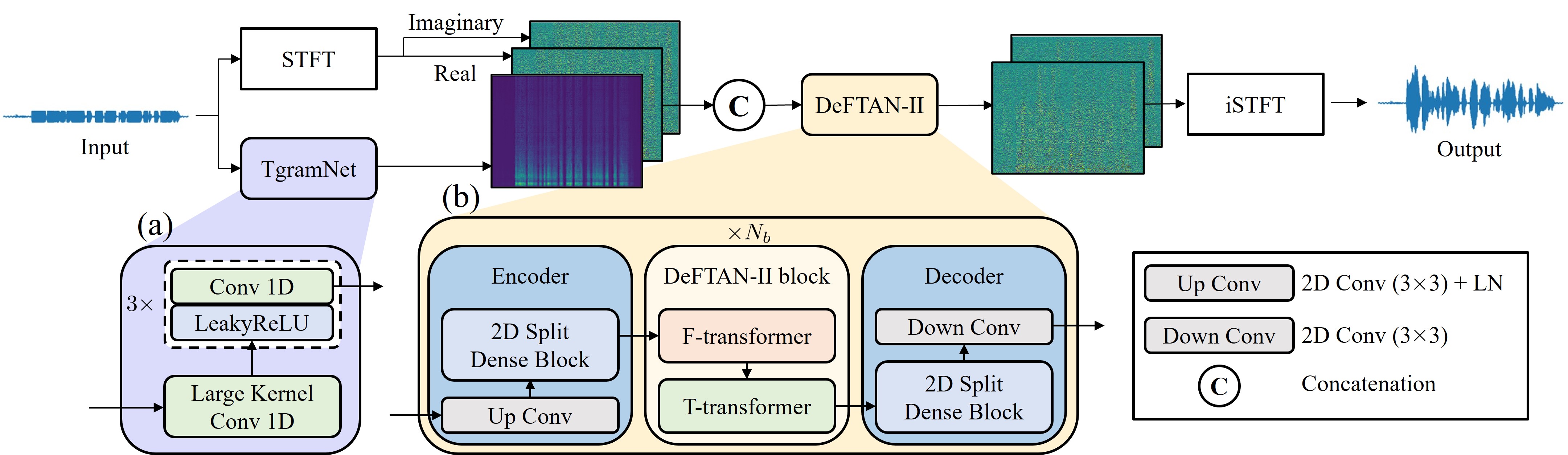}
    \caption{Overall architecture of the proposed declipping model (a) TgramNet (b) DeFTAN-II}
    \label{fig:overall_architecture}
\end{figure*}

Demucs-Discriminator-Declipper (DDD)~\cite{ddd} was introduced to reduce inference latency and enhance performance on low-SDR inputs by combining a speech enhancement model with adversarial training. The speech enhancement model, Demucs~\cite{demucs}, serves as the generator, while the HiFiGAN~\cite{hifigan} discriminator is incorporated to further improve perceived speech quality. As a fully DNN-based declipper, DDD performed well across a wide range of SDRs and achieved high scores in listening tests. 
However, artifacts such as the perception of missing fundamental frequencies and vowel distortion were observed in declipped samples.

The success of DDD highlights that well-designed speech enhancement models can effectively function as speech declipping models. In recent years, speech enhancement techniques have advanced from U-Net to transformer-based models~\cite{DPTNet, TFPSNet, TFGridNet, deftan, deftanii}. Transformer-based models capture temporal and spectral relationships in speech data on local and global scales through their multihead self-attention mechanisms, demonstrating strong performance even in extremely noisy and reverberant conditions. Especially, the critical importance of capturing spectral information has been emphasized in ~\cite{complexspectralmapping}, leading to the development of models with frequency transformers~\cite{TFPSNet, TFGridNet, deftan} that operate on complex spectrogram data.

In this work, we leverage the advantages of recent transformer-based speech enhancement models. However, complex spectrogram-based approaches would struggle to maintain consistency with reliable samples in the time domain and exhibit weakness in handling high-SDR inputs. 
To address this challenge, we incorporate a learnable temporal feature as the input to the transformer model, inspired by similar issues in anomalous sound detection (ASD). A recent ASD model, STgram-MFN~\cite{tgram}, demonstrated that spectrogram-based input features alone were insufficient for capturing key characteristics of machine sounds. This limitation was addressed by concatenating spectrogram features with temporal features extracted directly from an input waveform by an auxiliary network, TgramNet. 
We adopt this approach by integrating TgramNet into the transformer-based speech enhancement model, DeFTAN-II~\cite{deftanii}, to provide additional temporal information and enhance the model’s adaptability to high-SDR inputs. 
Our results show that this combination outperforms conventional models by a significant margin in the perceptual evaluation of speech quality (PESQ), Deep noise suppression mean opinion score (DNSMOS), and SDR on the clipped part of the waveform ($\text{SDR}_\text{c}$). 

\section{Declipping Architecture Design}
\subsection{Problem Statement}
A speech signal $\mathbf{y}\in \mathbb{R}^N$ clipped by the threshold $\theta$ from the clean speech signal $\mathbf{x}\in \mathbb{R}^N$ can be represented as 
\begin{equation}
    \mathbf{y}[n]=\begin{cases}
        \mathbf{x}[n] & \text{if}~|\mathbf{x}[n]|\leq \theta \\
        \theta \cdot \text{sign}({\mathbf{x}[n]}) & \text{otherwise}
    \end{cases},
\end{equation}
where $n=0,1,\cdots,N-1$ is the time index for the speech signals of length $N$. The objective of speech declipping is to restore $\hat{\mathbf{x}}$, a speech signal similar to $\mathbf{x}$, from $\mathbf{y}$. 

\subsection{Declipping Architecture}
The overall architecture of the proposed transformer-based declipping model is presented in Fig. \ref{fig:overall_architecture}. 
The input to the model is the clipped waveform $\mathbf{y}$, which is firstly transformed into a complex spectrogram $\mathbf{Y}\in\mathbb{R}^{2\times F\times T}$ using STFT, where $F$ and $T$ denote the number of frequency bins and time frames, respectively. 

Simultaneously, the input waveform is also processed by TgramNet to extract auxiliary temporal features $\mathbf{Y}_{T}\in\mathbb{R}^{1\times F\times T}$. 
The vanilla TgramNet proposed in \cite{tgram} consists of a single 1D conv with large kernels, followed by three repetitions of LayerNorm, leakyReLU, and 1D conv layers.
However, the use of LayerNorm restricts the flexibility in the number of time frames, so we omitted the LayerNorm layers, making TgramNet adaptable to varying input waveform lengths.
The initial 1D conv of TgramNet employs $F$ large kernels with kernel sizes and strides corresponding to the STFT window length and hop size, respectively, to convert the input waveform into a tensor with the same width and height as $\mathbf{Y}$. The following three 1D conv layers preserve the feature dimensions.
The complex spectrogram and temporal features are then concatenated along the channel dimension, resulting in the combined input tensor $\mathbf{Y}_{in}\in\mathbb{R}^{3\times F\times T}$.

The concatenated input tensor is then processed by the DeFTAN-II model~\cite{deftanii}, which comprises an encoder, a decoder, and $N_b$ transformer blocks (DeFTAN-II blocks) between them.
The first convolution layer of the encoder (Up Conv) is 2-D convolutional layer extracting local TF features and increasing the number of features from $3$ to $64$. This is followed by a 2D Split-Dense Block (SDB) further analyzing these features through a series of groupwise convolution and concatenation.
Next, DeFTAN-II blocks, consisting of the F- and T-transformers, capture relationships between frequencies and temporal frames and dynamically adapt to varying signal's characteristics using its self-attention mechanism. The output features from the final DeFTAN-II block are then aggregated by another 2D SDB and processed by a 2D transposed convolution layer (Down Conv) in the decoder, reconstructing the RI components of a declipped speech. From the reconstructed complex spectrogram, the inverse STFT outputs the declipped waveform  $\hat{\mathbf{x}}$.

\subsection{Training Objective}

The proposed network is trained by the weighted sum of the L1 loss of waveform ($\mathcal{L}_1$) and the multi-resolution STFT (MRSTFT) loss ($\mathcal{L}_S$), defined as
\begin{align}
\begin{split}
    &\mathcal{L}(\mathbf{x},\hat{\mathbf{x}})=\lambda_1\mathcal{L}_1 + \lambda_2\mathcal{L}_S 
    \\
    &= \lambda_1\frac{1}{N}\lVert\mathbf{x}-\hat{\mathbf{x}}\rVert_1 +\lambda_2\sum_{i=1}^{3}\big(\mathcal{L}_{SC}^{(i)}(\mathbf{x},\hat{\mathbf{x}})+\mathcal{L}_{mag}^{(i)}(\mathbf{x},\hat{\mathbf{x}})\big),
    \label{eqn:total_loss}
\end{split}
\end{align}
where $\lambda_1=100$ and $\lambda_2=1$, emphasizing the waveform L1 loss to direct more gradient flow carrying time-domain information to the STFT-based model. 
The MRSTFT loss consists of spectral convergence ($\mathcal{L}_{sc}$) and magnitude ($\mathcal{L}_{mag}$) losses, calculated across three STFT configurations using different FFT points [512, 1024, 2048] and hop sizes [50, 120, 240], respectively:  
\begin{align}
    \mathcal{L}_{\text{sc}}^{(i)}(\mathbf{x},\hat{\mathbf{x}})&=\frac{\lVert|\text{STFT}^{(i)}(\mathbf{x})|-|\text{STFT}^{(i)}(\hat{\mathbf{x}})|\rVert_2}{\lVert|\text{STFT}^{(i)}(\mathbf{x})|\rVert_2}\\
    \mathcal{L}_{\text{mag}}^{(i)}(\mathbf{x},\hat{\mathbf{x}})&=\lVert\log|\text{STFT}^{(i)}(\mathbf{x})|-\log|\text{STFT}^{(i)}(\hat{\mathbf{x}})|\rVert_{1}
\end{align}

\begin{table*}[t]
\caption{Performance comparison on VoiceBank-DEMAND and DNS Challenge testset}
\centering
\vspace{-1em}
\setlength{\tabcolsep}{5pt}
\renewcommand{\arraystretch}{1.2}
\resizebox{1\textwidth}{!}
{
\begin{tabular}{c|c|ccccc|ccccc|ccccc|cccc}
\hline
 & & \multicolumn{19}{c}{Input SDR (dB)} \\ 
\cline{3-21}
 & & ~1 & ~3 & ~7 & ~15 & INF & ~1 & ~3 & ~7 & ~15 & INF & ~1 & ~3 & ~7 & ~15 & INF & ~1 & ~3 & ~7 & ~15 \\
\cline{3-21}
 Datasets & Methods & \multicolumn{5}{c|}{PESQ $\uparrow$} & \multicolumn{5}{c|}{DNSMOS $\uparrow$} & \multicolumn{5}{c|}{SDR (dB) $\uparrow$} & \multicolumn{4}{c}{$\text{SDR}_\text{c}$ (dB) $\uparrow$} \\
\hline
\hline
& Clipped   & 1.15 & 1.39 & 2.05 & 3.24 & 4.50
 & 2.75 & 3.01 & 3.25 & 3.45 & \underline{3.53}
 & ~1.00 & ~3.00 & ~7.00 & 15.00 & INF
 & ~0.90 & ~2.66 & ~5.78 & 10.70 \\
& A-SPADE~\cite{sparse_2015_aspade}  & 1.54 & 2.02 & 2.84 & 3.89 & -
 & 2.94 & 3.10 & 3.26 & 3.44 & - 
 & ~5.79 & ~7.73 & 12.58 & 21.36 & -
 & ~5.89 & ~7.13 & 10.75 & 15.94 \\
& T-UNet~\cite{tunet}   & 2.80 & 3.42 & 3.91 & 4.31 & 4.50
 & 3.11 & 3.23 & 3.39 & 3.49 & 3.52 
 & ~9.56 & 13.83 & 18.40 & 25.07 & 31.32
 & ~9.68 & 13.87 & 17.66 & 22.10 \\
VBDM & DD        & 2.92 & 3.44 & 3.94 & \underline{4.35} & 4.50
 & 3.13 & 3.23 & 3.40 & 3.50 & \underline{3.53} 
 & 11.99 & \underline{14.84} & 18.83 & \textbf{25.81} & 32.06
 & 12.20 & \underline{14.74} & 17.68 & 21.52 \\
& DDD~\cite{ddd}      & 2.45 & 3.31 & 3.98 & 4.34 & 4.50
 & \textbf{3.32} & 3.36 & 3.42 & 3.49 & 3.52 
 & ~9.72 & 13.55 & 17.87 & 23.59 & 29.71
 & 10.09 & 13.85 & 17.58 & 20.64 \\
\cline{2-21}
& DeFTAN-II~\cite{deftanii} & \underline{3.26} & \underline{3.68} & \underline{4.00} & 4.16 & 4.24
 & 3.24 & \underline{3.38} & \underline{3.48} & \underline{3.51} & 3.52 
 & \textbf{12.78} & \textbf{15.43} & \underline{19.20} & 24.98 & \underline{30.71}
 & \textbf{13.19} & \textbf{15.84} & \underline{19.08} & \underline{22.61} \\
& Proposed  & \textbf{3.30} & \textbf{3.78} & \textbf{4.20} & \textbf{4.44} & \textbf{4.50}
 & \underline{3.25} & \textbf{3.43} & \textbf{3.52} & \textbf{3.53} & \textbf{3.54} 
 & \underline{12.70} & \textbf{15.43} & \textbf{19.44} & \underline{25.34} & \textbf{39.63}
 & \underline{13.09} & \textbf{15.84} & \textbf{19.35} & \textbf{23.05} \\ 
\hline \hline
& Clipped   & 1.14 & 1.33 & 1.93 & 3.31 & \textbf{4.50} 
 & 2.47 & 2.87 & 3.42 & 3.83 & \textbf{4.02} 
 & ~1.00~ & ~3.00 & ~7.00 & 15.00 & INF 
 & ~0.89 & ~2.53 & ~5.24 & ~9.19 \\
& A-SPADE~\cite{sparse_2015_aspade}  & 1.37 & 1.73 & 2.66 & 3.98 & -    
 & 3.14 & 3.33 & 3.61 & 3.90 & - 
 & ~4.71 & ~5.86 & 10.70 & 22.09 & - 
 & ~5.09 & ~8.01 & 11.93 & 17.68 \\
& T-UNet~\cite{tunet}   & 1.99 & 2.80 & 3.57 & 4.22 & 4.43 
 & 3.21 & 3.50 & 3.75 & 3.95 & 3.99 
 & ~5.31 & ~\textbf{9.46} & \textbf{14.49} & \textbf{22.15} & 27.97 
 & ~5.73 & ~\underline{9.90} & \underline{13.86} & \underline{18.70} \\
DNS & DD        & \underline{2.12} & 2.79 & 3.58 & 4.28 & \textbf{4.50} 
 & 3.28 & 3.51 & 3.78 & 3.96 & \underline{4.01} 
 & ~6.34 & ~\underline{9.27} & \underline{13.81} & \underline{21.86} & 28.00 
 & ~6.78 & ~9.67 & 13.10 & 17.52 \\
& DDD~\cite{ddd}      & 1.78 & 2.60 & 3.53 & 4.25 & 4.50
 & 3.50 & 3.72 & 3.84 & 3.95 & \underline{4.01} 
 & ~5.17 & ~8.57 & 12.66 & 20.14 & 29.71
 & ~5.50 & ~9.06 & 12.55 & 16.90 \\
\cline{2-21}
& DeFTAN-II~\cite{deftanii} & \textbf{2.54} & \underline{3.12} & \underline{3.76} & \underline{4.30} & 4.43 
 & \textbf{3.71} & \underline{3.84} & \underline{3.91} & \underline{3.97} & 4.00 
 & ~\textbf{6.75} & ~9.06 & 13.00 & 20.76 & \underline{37.74} 
 & ~\underline{7.29} & ~9.89 & 13.31 & 17.97 \\
& Proposed  & \textbf{2.54} & \textbf{3.14} & \textbf{3.81} & \textbf{4.37} & \textbf{4.50} 
 & \underline{3.66} & \textbf{3.85} & \textbf{3.96} & \textbf{4.00} & \textbf{4.02} 
 & ~\underline{6.63} & ~9.01 & 13.23 & 20.74 & \textbf{38.32} 
 & ~\textbf{7.33} & \textbf{10.05} & \textbf{13.90} & \textbf{18.94} \\ 
\hline
\end{tabular}
}
\label{VBDM_result}
\end{table*}

\section{Experiments}
\subsection{Dataset \& Preprocessing}

The training and test datasets were constructed by following the setup similar to that of~\cite{ddd}. The clean data for training were obtained from the clean speeches of the VoiceBank-DEMAND (VBDM) dataset~\cite{valentini}, which includes 9.4 hours of speech from 28 speakers. However, we selected the clipping threshold $\theta$ from a uniform distribution between minimum and maximum values of [0.01, 0.125] (1 dB to 9 dB SDRs), instead of the sampling in the exponential scale of~\cite{ddd}. This uniform sampling in a linear scale helped the DNN adapt better over the wide range of SDRs. 

For the testing, we prepared two datasets: (1) VBDM data unseen during the training and (2) the deep noise suppression challenges (DNS) dataset~\cite{dns} to evaluate adaptability to the unseen corpora.
The VBDM test dataset contains 35 minutes of speech from two speakers of different genders, while the DNS test dataset comprises 150 samples, each 10 seconds long, for a total of 25 minutes. 
Each dataset was clipped at SDR levels of 1 dB, 3 dB, 7 dB, and 15 dB, with unclipped data also included in the evaluation. 
For both training and testing, all signals were downsampled to 16 kHz.

\subsection{Experimental Settings}
For performance comparison, we selected A-SPADE and three DNN networks, T-UNet, DD, and DDD as baseline models. The models were implemented following the architectures described in their respective original studies. Along with the proposed model, we also considered the DeFTAN-II model without TgramNet to investigate the efficacy of learned temporal features. 
FOr training models, AdamW optimizer was used, and the learning rate for the baseline models was the same ($10^{-4}$) as that used for their original studies while $10^{-3}$ was used for the proposed model. 
The DDD model with adversarial training was trained with a batch size of $2$, while the other models were trained with a batch size of $8$.
All models were trained for up to 150 epochs, with validation performed at the end of each epoch. The test was conducted on the models with the lowest validation loss.

\subsection{Objective Metrics}
The declipping performance was evaluated using four metrics: PESQ, DNSMOS, SDR, and $\text{SDR}_\mathrm{c}$. 
PESQ assesses the perceptual quality of speech by comparing a degraded speech to a reference signal, with scores ranging from -0.5 to 4.5.
Higher PESQ scores indicate a closer perceptual similarity to the reference speech.
DNSMOS evaluates the perceptual quality of denoised speech through a DNN model trained to estimate the mean opinion score (MOS). In principle, the declipping task is different from the denoising task, but DNSMOS was employed as an indirect indicator of audible distortions. 
SDR measures the energy ratio of the reference signal to the distortion in the reconstructed signal, qualifying how numerically close the declipped speech is to the reference.
Our primary interest lies in the distortion occuring in the clipped region, so we conducted a separate evaluation using $\text{SDR}_{\text{c}}$~\cite{servey_article}, which specifically measures the SDR in the clipped region.

\section{Results}

\subsection{Objective Evaluations}
Table \ref{VBDM_result} presents the results evaluated on the VBDM and DNS test dataset.
The column labeled `INF' represents the performance on unclipped speech inputs. Boldface and underlined numbers indicate the best and the second-best results for each metric, respectively.

The first result is from the test on the VBDM dataset, which includes the unseen speech instances from the same speech corpora as the training dataset. 
For PESQ, the two transformer-based models (DeFTAN-II and Proposed) achieved scores exceeding 3.26 for 1 dB SDR input, significantly outperforming the highest baseline score of 2.92.
However, vanilla DeFTAN-II exhibited weaknesses with the high SDR inputs, yielding lower PESQ scores than other baselines for 15 dB SDR and unclipped inputs. This limitation was addressed in the proposed architecture with TgramNet, achieving PESQ scores of 4.44 for the 15 dB SDR and 4.50 for unclipped speech. Notably, the score for the 15 dB SDR input surpassed those of all baselines, and the score for unclipped speech reached the maximum PESQ value.
Overall, the proposed model consistently outperformed baselines in PESQ across all input SDRs.

The previous study~\cite{ddd} reported that DDD achieved high subjective test scores by improving the perceptual speech quality rather than objective metrics. In our tests using DNSMOS, DDD exhibited a similar trend, showing a high DNSMOS score for the 1 dB SDR input. 
However, except for this case, the proposed model constantly achieved the highest DNSMOS scores compared to other baselines, including the vanilla DeFTAN-II. This further emphasizes that the integration with TgramNet improves the perceptual quality of the declipped speech.

In terms of SDR improvement, the transformer-based models increased SDRs by 0.79 dB and 0.71 dB for 1 dB SDR input compared to the highest-performing baseline model (DD). These models also outperformed the baselines at 3 dB and 7 dB SDR inputs. The only exceptional case is for the 15 dB input, where DD showed the highest score. However, in terms of $\text{SDR}_\mathrm{c}$, the proposed model's score is significantly higher than DD. This implies that the high SDR of DD is obtained from the unclipped part of the signal rather than the clipped part. 
For unclipped speech, the proposed model achieved the highest SDR of nearly 40 dB, which is 7 dB higher than any baseline.

In tests conducted on the DNS dataset, similar trends were observed as in the VBDM results. The two transformer-based models consistently delivered the best or second-best scores across most metrics and input SDRs. One notable difference was the higher SDRs of the DD model on the DNS dataset. However, the $\text{SDR}_\mathrm{c}$ of DD was lower than that of the proposed model, indicating that DD tends to focus more on preserving reliable samples compared to other models. Even when its SDR is lower than that of DD, the proposed or DeFTAN-II model demonstrated superior performance in PESQ, DNSMOS, and $\text{SDR}_\mathrm{c}$, highlighting their ability to better handle the clipped regions.

\subsection{Qualitative Analysis}\label{q_analysis}
\begin{figure}[t]
    \centering
    \includegraphics[width=8.5cm]{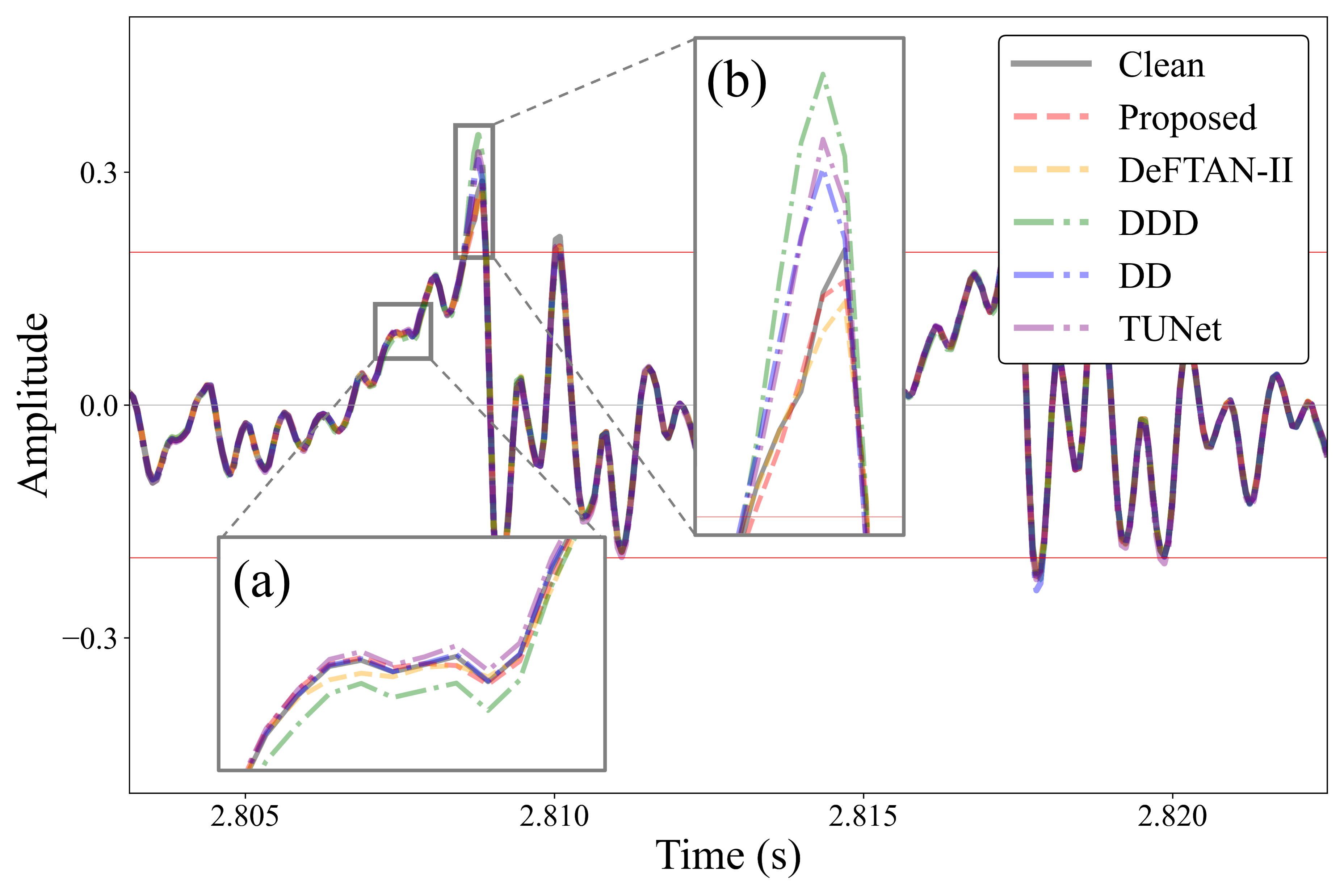}
    \vspace{-1em}
    \caption{Waveforms of clean and declipped speeches, processed by individual declipping models for a 15 dB SDR input signal. Declipped waveform in (a) the unclipped region and (b) the clipped region.}
    \label{fig:waveform_comparison}
\end{figure}

Fig. \ref{fig:waveform_comparison} compares the waveforms from the VBDM dataset, which were declipped from a speech signal clipped at an input SDR of 15 dB.
The red horizontal lines represent the clipping threshold, while the black curve indicates the reference clean speech.  
It is noteworthy that the transformer-based models restore the clipped sections above the threshold with shapes almost identical to the reference speech, whereas the baseline models often overestimate the clipped regions. This superior reconstruction by the transformer-based models explains their higher PESQ scores and $\text{SDR}_\mathrm{c}$s for the 15 dB SDR input. 
However, despite having lower reconstruction performance, DD and T-UNet exhibited a higher output SDR for this high-SDR input. This can be explained by time-domain models' ability to preserve the unclipped sections of the waveform almost perfectly, allowing it to achieve high SDRs by minimizing distortions in the unclipped regions, rather than focusing on reconstructing the clipped regions. This result suggests that while the transformer-based models excel in reconstructing clipped region, there is potential for further improvement. If the architecture can automatically replace the unclipped regions with those of the input signal, the overall performance, particularly in terms of SDR, could be enhanced even further. This remains a promising direction for future research. The declipped speech samples can be accessed at the website\footnote{\url{https://k0hoo.github.io/transformer-declipper/}}.

\section{Conclusion}

This work proposed a transformer-based declipping architecture that simultaneously utilizes complex spectrogram input and learnable temporal features. The transformer-based models demonstrated significant performance improvements over baseline models, particularly for low-SDR input signals. Moreover, the proposed model with learnable temporal features outperformed the vanilla transformer model relying solely on spectrogram input, showing marked improvement for high-SDR inputs.
The proposed architecture, evaluated for VBDM, delivered exceptional results across all metrics, including PESQ, DNSMOS, SDR, and $\text{SDR}_\mathrm{c}$. It also exhibited minimal distortion on clean inputs and unseen corpora of DNS datasets, demonstrating its adaptability and robustness across a wide range of SDR conditions.




\vfill\pagebreak

\vspace{12pt}

\end{document}